\documentclass[prb,aps,twocolumn,showpacs]{revtex4-1}

\usepackage{graphicx}% Include figure files
\usepackage{here}
\usepackage{bm,color}% bold math
\usepackage{physics}
\usepackage{amsmath}
\usepackage{bm,color}
\usepackage{multirow}

\newcommand{\be}{\begin{eqnarray}}
\newcommand{\ee}{\end{eqnarray}}

\usepackage{color}
\usepackage{ulem}

\def\r{\rangle}
\def\l{\langle}
\def\trho{\tilde{\rho}}
\def\lm{\lambda}
\def\A{\hat{x}^2_j+\hat{\lambda}_j^2}
\def\B{x^2_j+\lambda^2_j}

\begin{document}

\title{
Deformation of localized states and state transitions 
in systems of randomly hopping interacting fermions}
\date{\today}
\author{Takahiro Orito$^{1}$}
\author{Yoshihito Kuno$^{2}$}
\author{Ikuo Ichinose$^{3}$}

\affiliation{$^1$Graduate School of Advanced Science and Engineering, Hiroshima University, 739-8530, Japan}
\affiliation{$^2$Graduate School of Engineering Science, Akita University, Akita 010-8502, Japan}
\affiliation{$^3$Department of Applied Physics, Nagoya Institute of Technology, Nagoya, 466-8555, Japan}

\begin{abstract}
We numerically study the random-hopping fermions (the Cruetz ladder) with repulsion and 
investigate how the interactions deform localized eigenstates
by means of the one-particle-density matrix (OPDM).
The ground state exhibits resurgence of localization from the compact localized state
to strong-repulsion-induced localization.
On the other hand, excited states in the middle of the spectrum tend to extend 
by the repulsion.
The transition property obtained by numerical calculations of the OPDM is deeply understood 
by studying a solvable model in which local integrals of motion (LIOMs) are obtained
explicitly.
The present work clarifies the utility of the OPDM and also how compact-support LIOMs in
non-interacting limit are deformed by the repulsion.
\end{abstract}
\maketitle

%%%%%%%%%%%%%%%%%%%%%%%%%%%%%%%%%%%%%%%%%%%%%%%%%%%%%%%%%%%%%%%%%%%%%%%%%%%%%%

\section{Introduction}

Understanding universal mechanism of localization phenomena \cite{Anderson,Abrahams1979} is 
an important topic in condensed matter physics \cite{Nandkishore2015,Imbrie2017}. Localization breaks the eigenstate-thermalization-hypothesis (ETH) \cite{ETH1,ETH2}, 
that is, the prescriptions of conventional statistical mechanics cannot be applied to 
the localized systems. 
Also certain interacting many-body systems exhibit exotic behaviors of localization and 
entanglement properties, which are called many-body localization (MBL)
\cite{Altshuler1997,Gornyi,Basko2006,Bardarson,Serbyn2013,Huse2014}. 
Although the presence of localized phenomena have been reported in various interacting
many-body-systems \cite{Abanin2017,Alet,Abanin2019}, universal understanding of
the origin and detailed mechanism of MBL has not been obtained yet.
Especially, how interactions affect localization phenomena in various 
many-body systems remains an open question.

In this work, we investigate effects of interactions in a localized model 
with disorders, namely, random-hopping
Creutz ladder \cite{Creutz1999,Bermudez,Junemann,Shin2020,KOI2020,kuno2020,OKI2021}. 
This model has short-range entangled clusters, i.e., compact-localized states (CLS), 
which often appear in flat-band systems~\cite{Leykam2018}. 
There are many works on localization in various models with the CLS including the cluster spin,
diamond chain, etc \cite{Bahri,KOI2020,Danieli_1,Roy,Zurita,OKI2021,Tilleke,Khare}, and they give important insight into localization. 
The present work belongs to the category of these studies and intends to clarify specific aspects of localization,
i.e., deformation of localized states by interactions. 
That is, we introduce interactions, which can deform and destroy the CLS since the operators of the CLSs no longer commute the full interacting Hamiltonian, thus the many-body states constructed by the CLSs are no longer exact many-body 
eigenstates in the interacting system. 
In particular, as we increase the strength of the interactions, 
how the CLS-based localized states deform and what states, including an extended (ETH) state, emerge 
as an alternative to them is an important and interesting problem. 
In this work, we shall focus on these issues and obtain clear understanding.

In general, studying such a quantum many-body system is not an easy task.
However, one-particle density matrix (OPDM) \cite{Bera2015,Bera2017,Hopjan2020,OK2021} is a useful and
promising tool, in particular, it reveals certain
important aspects of localization emerging in disorder systems. 
In this work, we make extensive use of the OPDM and analyze the random-hopping Creutz ladder.

The rest of this paper is organized as follows.
In Sec. II, we shall introduce the random-hopping Creutz ladder with interactions.
The basic properties of the non-interacting system is explained.
We then introduce the OPDM, and its relation to the Fock-state view point of localization.
In Sec. III, we present results of the numerical study by means of the exact diagonalization.
Not only the distribution of eigenvalues of the OPDM, but also various quantities 
derived from the OPDM, e.g.,
the distribution of the diagonal elements of the OPDM, the real part of the OPDM elements
in the real-space and single-particle bases, etc, are obtained. 
These results clarify how the ground state (GS) and spectrum-middle states change as the strength of the
interactions is increased and how the CLS picture breaks down.
In Sec. IV, we examine the observations obtained by the numerical methods by considering solvable models.
Details of the transition of the states are clarified there.
Section V is devoted for discussion and conclusion.

%%%%%%%%%%%%%%%%%%%%%%%%%%%%%%%%%%%%%%%%%%%%%%%%
\section{Random-hopping Creutz ladder and OPDM}

We first introduce the target model; random-hopping
Creutz ladder, whose Hamiltonian is given as, 
\be
H_{\cal T} &=& H_{\rm rh}+H_{\rm I},  \nonumber \\
H_{\rm rh} &=& -\sum^{L}_{j=1}t_j [ia^\dagger_{j+1}a_j-ib^\dagger_{j+1}b_j
+a^\dagger_{j+1}b_j+b^\dagger_{j+1}a_j]  \nonumber \\
&&+\mbox{h.c.},   \label{HT}
\ee
\be
H_{\rm I} &=& V \sum^{L}_{j=1}\biggl[\biggl(n^a_{j}-\frac{1}{2}\biggr)\biggl(n^a_{j+1}-\frac{1}{2}\biggr)\label{HI1}\\
&&+\biggl(n^b_j-\frac{1}{2}\biggr)\biggl(n^b_{j+1}-\frac{1}{2}\biggr)+\biggl(n^a_j-\frac{1}{2}\biggr)\biggl(n^b_j-\frac{1}{2}\biggr)\biggl],\nonumber
%\label{HT}
\ee
where $t_j=1+\delta_j$ is the amplitude of the random hopping and we set 
$\delta_j \in [-0.3.+0.3]$ (uniform distribution) and $n^{a}_{j}=a^{\dagger}_j a_j$, etc. 
The parameter $V(>0)$ is the strength of the repulsion and a varying parameter.
In this  work, we mostly study the GS and states in the middle of 
the spectrum at half-filling.
The reason why we take some specific hopping amplitudes in $H_{\rm rh}$ [Eq.~(\ref{HT})] 
gets clear
by introducing two new fermionic operators defined on {\it each plaquette}, 
\be
&& W^+_j={1 \over 2}(-ia_{j+1}+b_{j+1}+a_j-ib_j), \nonumber \\
&& W^-_j={1 \over 2}(-ia_{j+1}+b_{j+1}-a_j+ib_j).
\label{WW}
\ee
Then in terms of $\{W^\pm_j\}$, the free Hamiltonian $H_{\rm rh}$ is expressed as, 
\be
H_{\rm rh} = -\sum^{L}_{j=1} 2t_j[W^{+\dagger}_j W^+_j-W^{-\dagger}_jW^-_j].
\label{Hrh}
\ee
As $\{W^{\pm}_j\}$'s are fermion operators satisfying canonical anti-commutators such as 
$\{W^{+\dagger}_i, W^+_j\}=\delta_{ij}$, etc, exact energy eigenstates of the non-interacting system, $H_{\rm rh}$,
are given by the Fock states of $\{W^\pm_j\}$, i.e., 
$
\prod_{j=1}^L[(W^+_j)^\dagger]^{m_j}[(W^-_j)^\dagger]^{\ell_j}|0\rangle,
\label{Fstate}
$
where $m_j=0, 1$ $\ell_j=0, 1$, and $\sum_{j=1}^L(m_j+\ell_j)=N$ for $N$-particle states. 
In particular for the GS at half-filling, the many-body state is given by
$|{\rm GS}_0\r=\prod^L_{j=1}[W^{+\dagger}_j]|0\r$.
On the other hand in the middle of the spectrum (zero energy), 
$|\psi\rangle \sim \prod_j [W^{+\dagger}_jW^{-\dagger}_j]|0\r$,
where $\{j\}$'s are rather arbitrary sites.

We rename the fermion operators at site in such a way; $(a_1,b_1,a_2, \cdots,b_L)=(c_1,c_2,c_3,\cdots,c_{2L})$, and then
OPDM, $\hat{\rho}=\{\rho_{\ell m}\}$, is defined as follows for a general (interacting) many-body state $|\psi\r$
using operators $\{c_\ell\}_{\ell=1}^{2L}$, which we call OPDM in the real-space basis,
\be
\rho_{\ell m}=\l\psi|c^\dagger_\ell c_m|\psi\r, \:\: \ell,m=1,2,\cdots,2L.
\label{rho1}
\ee
By diagonalizing $\hat{\rho}$, non-negative eigenvalues 
$\{n^\alpha\}$ ($\alpha=1, \cdots, 2L$) and eigenfunctions $\{\phi^\alpha\}$
of $\hat{\rho}$ are obtained.
For the system of total particle number $N$, one can show $\sum^{2L}_{\alpha=1} n^\alpha=N$.
In MBL regimes, $n^\alpha\sim 1$ or $0$, and the eigenfunctions (orbitals) corresponding to $n^\alpha\sim 1$
are regarded as occupied single-particle states and those for $n^\alpha\sim 0$ empty states.
In this case, the state,  $|\psi\r$, is well-described by a Slater determinant (SD) of
$\{\phi^\alpha\}$ for $n^\alpha\sim1$.

In the subsequent sections, we shall numerically study how the states in $H_{\rm rh}$ changes by introducing 
the interactions $H_{\rm I}$ by observing the OPDM.
To this end, we define $W$-representation of OPDM such as
$\l W^\dagger_{\ell} W_m\r$, explicit form of which is defined later on.
In the previous studies, the properties of the OPDM, in particular its eigenvalues, were employed for the
study of localization.
In the present work, we shall make full use of the OPDM in order to clarify how states change
under the increase in the strength of the interactions.
This is one of the main highights of the present work as we explained in Introduction.

It is quite instructive and useful to examine the OPDM 
from the view point of Fock-space localization~\cite{Logan2019,Tomasi2020,OK2021,Roy2021}.
The elements of the OPDM are nothing but correlations on the Fock-space for an eigenstate.
To express many-body states with $L$ sites, 
we introduce suitable orbitals $\{d_j\}$, where index $j$ does not 
necessarily represent lattice site although we use the same notation.
Then, the basis of the Fock-space is given by 
$|I\rangle=\prod_{j=1}^L[d_j^\dagger]^{m_j}|0\rangle$,
where $m_j=0, 1$ and $\sum_{j=1}^Lm_j=N$. 
Many-body state is generally expressed as
$|E\rangle=\sum_I A_I|I\rangle$ where $\sum_I|A_I|^2=1$.

Here, we introduce Hamming distance between two Fock-states $|I\r$ and $|K\r$, 
defined as 
$r_{IK}={1\over 2}\sum_{j=1}^L(m_{j,I}-m_{j,K})^2$,
with 
$N_j\equiv d_j^\dagger d_j$, $N_j|I\rangle=m_{j,I}|I\rangle, \ N_j|K\rangle=m_{j,K}|K\rangle$.
Then, correlation function on the Fock-space is defined as 
\be
F(r)=\sum_{I,K:,r_{IK}=r}|A_I|^2|A_K|^2.
\ee
The correlation function can quantify the localization tendency of the state.
In a localized state $|E\r$, there are $N$ orbitals (eigenvectors of the OPDM) with eigenvalue 
$n^\alpha\sim1 \ (\alpha=1, \cdots, N)$,
and there is a dominant Fock state $|M_0\r$, which is close to the SD of 
these ${\phi^\alpha} \ (\alpha=1, \cdots, N)$.
Then, an eigenstate $|E\r$ is given as
\be
|E\rangle=A_{M}|M_0\rangle+A_K|K\rangle+\cdots,\nonumber
%\label{state1}
\ee
Let us assume $d^\dagger_id_j|M_0\rangle=|K\rangle \ (i\neq j)$, and then the off-diagonal
($i\neq j$) element of the OPDM, $\rho_{ij}=\langle E|d^{\dagger}_{i}d_{j}|E\rangle$ 
is given by
\be
&& [\langle M_0|A^\ast_M+\langle K|A^\ast_K+\cdots] d^\dagger_i d_j
[A_{M}|M_0\rangle+A_K|K\rangle+\cdots]  \nonumber \\
&& \sim A^\ast_KA_M.
\label{OPDM1}
\ee
Eq.~(\ref{OPDM1}) shows that the distance between
$|M_0\rangle$ and $|K\rangle$ is $r_{M_0K}=2$, and therefore,
$|\rho_{ij}|^2=|A_MA_K|^2$ measures mixing amplitude of the leading state $|M_0\rangle$
and the states with distance $r_{M_0K}=2$.
One may expect that the off-diagonal elements of the OPDM are getting large at a phase
transition of MBL as the localized state dominated by the SD, $|M_0\r$, is destroyed by
the growing subleading states such as $|K\r$.
Detailed discussion on this point will be given later on by studying the
Creutz ladder.

%%%%%%%%%%%%%%%%%%%%%%%%%%%%%%%%%%%% 
\begin{figure}[t]
\begin{center} 
\includegraphics[width=8.5cm]{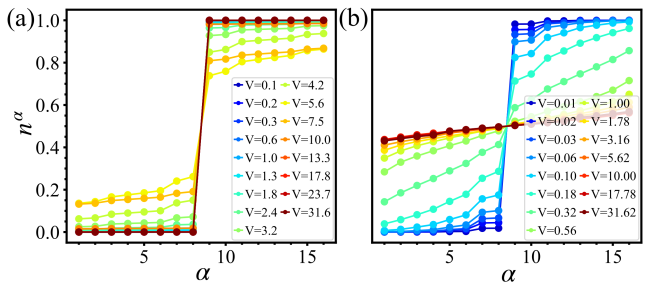}  
\end{center} 
%\vspace{-0.5cm}
\caption{
Eigenvalues of the OPDM for the GS (a) and the band center (b). 
$\alpha$ is the number of eigenstate of the OPDM of Eq.~(\ref{rho1}) 
in ascendant energy order.
Results are averages $5000$ samples of the eigenstates. 
}
\label{Fig1}
\end{figure}
%%%%%%%%%%%%%%%%%%%%%%%%%%%%%%%%%%%%

%%%%%%%%%%%%%%%%%%%%%%%%%%%%%%%%%%%%%%%%%%%%%%%%%%%
\section{Numerical studies}

We move on the numerical study of the Creutz ladder.
We mostly set the system size $2L=16$ and particle number $N=8$ \cite{Quspin}.

In Fig.~\ref{Fig1}, we first show how the eigenvalues of the OPDM, $\{n^\alpha\}$, vary
as the repulsion, $V$, increases. 
In the case of small $V$,
for both the GS and band center, $n^\alpha \sim 1$ for $1\leq \alpha \leq N=L$, and 
$n^\alpha \sim 0$ for $N < \alpha$, indicating localization takes place there.
As $V$ increases, the GS keeps this behavior of $\{n^\alpha\}$ until the step-wise shape
gets weaker at intermediate $V$'s, and for further increase of $V$, $\{n^\alpha\}$ resume the behavior for small $V$.
On the other hand in the band center, almost all eivenvalues $\{n^\alpha\} \sim 0.5$ for $V\geq 0.50$, 
indicating breakdown of the states composed of $|\psi\rangle \sim \prod_j  [W^{+\dagger}_jW^{-\dagger}_j]|0\r$.

%%%%%%%%%%%%%%%%%%%%%%%%%%%%%%%%%%%% 
\begin{figure}[t]
\begin{center} 
\includegraphics[width=8.5cm]{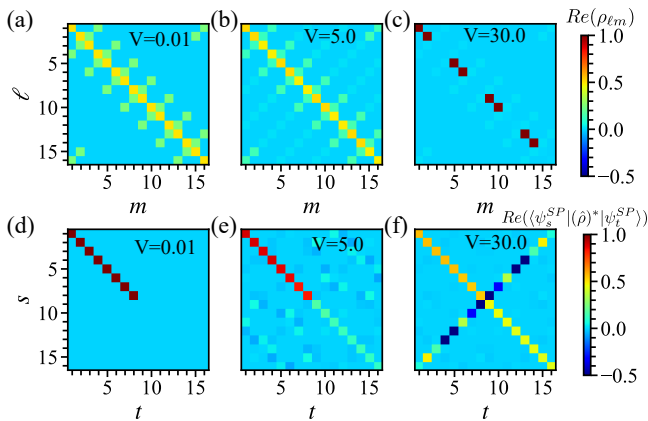}  
\end{center} 
%\vspace{-0.5cm}
\caption{
(a), (b) and (c): Real part of the OPDM of the GS for $V=0.01, 5.00$ and $30.0$.
Figures represent the OPDM of Eq.~(\ref{rho1}) in the real space such as $(c_1,c_2,c_3,\cdots,c_{2L})$.
(d), (e) and (f): Real part of $\langle \psi^{SP}_{s}|(\hat{\rho})^{*}|\psi^{SP}_{t}\rangle$. 
The state labels $s$ and $t$ are in the ascendant energy order.
Each result is obtained from single shot for single disorder realization.
}
\label{Fig2}
\end{figure}
%%%%%%%%%%%%%%%%%%%%%%%%%%%%%%%%%%%%
%%%%%%%%%%%%%%%%%%%%%%%%%%%%%%%%%%% 
\begin{figure}[t]
\begin{center} 
\includegraphics[width=5.5cm]{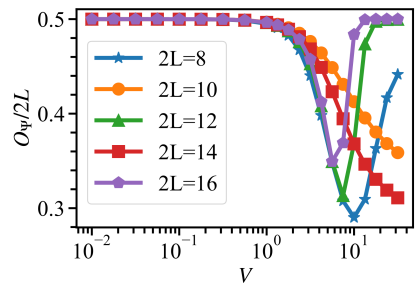}  
\end{center} 
%\vspace{-0.5cm}
\caption{
$O_\Psi=\langle\mbox{Tr}  [\hat{\rho}^2]\rangle$ of the GS at half-fillings. 
For $2L=8, 12$ and $16$, $O_\Psi$ first exhibits a sharp decrease and then, 
it returns to the original value quite rapidly as $V$ increases.
This behavior indicates that a GS transition to a novel localized state takes place.
On the other hand for $2L=10$ and $14$, this recurrence of $O_\Psi$ does not occur.
These results indicate that a state with the rung density wave emerges
at the commensurate fillings for $V>V_c$ as Fig.~\ref{Fig2}(c) implies.
}
\label{Fig3}
\end{figure}
%%%%%%%%%%%%%%%%%%%%%%%%%%%%%%%%%%%%

In order to clarify the above observed behavior of the states, 
we calculate the real part of 
the OPDM of the GS in the real space of $(a_1,b_1,a_2,b_2, \cdots, a_L,b_L)$
[$\{\rho_{\ell m}\}$ of Eq.~(\ref{rho1})] and obtain the results as in Figs.~\ref{Fig2}(a), \ref{Fig2}(b) and \ref{Fig2}(c).
In these data, as $V$ increases, it is obvious that the GS changes from the $W^\pm$-dominant state to 
the real space $ab$-dominant one.
For large $V$, the GS has a density-wave order as shown in Fig.~\ref{Fig2}(c). 
Additional data of the OPDM for in critical regime ($V\sim 10$) and its density
distribution are displayed in Appendix A.

We also calculate OPDM in the $W$-state basis ($W$-basis OPDM):  
$|\psi^{SP}_{\pm j}\rangle=(W^{\pm}_{j})^{\dagger}|0\rangle$ ($j=1,2,\cdots,L$), 
where $|\psi^{SP}_{\pm j}\rangle$ is obtained from the definition of $W^\pm_j$ as an linear combination of
the Fock-state  basis $\{c^\dagger_\ell|0\rangle\}$.
Then, the $W$-basis OPDM is defined by 
$$
\langle \psi^{SP}_{\pm i}|(\hat{\rho})^{*}|\psi^{SP}_{\pm j}\rangle 
\equiv \sum_{\ell, m}
\langle \psi^{\rm SP}_{\pm i} | c^\dagger_\ell |0\rangle \rho^\ast_{\ell m}\langle 0|c_m|\psi^{\rm SP}_{\pm j}\rangle ,
$$
where the matrix $(\hat{\rho})^{*}$ is the complex conjugate of 
$\hat{\rho}$. 
[The necessity of taking the complex conjugate is clarified by the
practical calculation.] 
Furthermore, we simplify the state notation of $|\psi^{SP}_{\pm j}\rangle$ as 
$(({+},{1}),({+},2),({+},3),\cdots,({+,L}), ({-,L}),\cdots, ({-,2}),({-,1})) \\
\to (1,2,\cdots,2L)\to (s=1,2,\cdots,2L)$ 
and then write the matrix element as $\langle \psi^{SP}_{s}|(\hat{\rho})^{*}|\psi^{SP}_{t}\rangle$ ($s,t=1,2,\cdots,2L$). 
In Figs.~\ref{Fig2}(d), ~\ref{Fig2}(e) and ~\ref{Fig2}(f), the real part of 
each element of the $W$-basis OPDM is shown.
In these data, while for small $V$, each orbital is identical to the state $(W^{\pm}_{j})^{\dagger}|0\rangle$, that is, the GS is a many-body state mainly 
described by a product of the $W$-basis, $(W^{+}_{j})^{\dagger}|0\rangle$.
As $V$ increases, this picture fades away and for large $V$, a different type of 
orbitals emerge, which are mixing state of $(W^{+}_{j})^{\dagger}|0\rangle$ and 
$(W^{-}_{j})^{\dagger}|0\rangle$ as we discuss later on. 
Thus, in the GS, the $W^{\pm}_{j}$ picture is significantly deformed
by the interactions.

To further quantify the above change of the OPDM, 
we introduce a novel quantity $O_\Psi=\langle\mbox{Tr}  [\hat{\rho}^2]\rangle$, 
where $\langle \cdots \rangle$ denotes the average over the disorder realization,
and $O_\Psi$ measures the magnitude of the diagonal elements. 
Numerical result of $O_\Psi$ for the GS
is displayed in Fig.~\ref{Fig3} indicating 
that a `phase transition' takes place at $V_c\sim 7$,
and a novel state emerges for $V>V_c$.
Interestingly enough, $O_\Psi$ behaves quite differently for the commensurate
($2L=8,12, 16$) and the incommensurate ($2L=10, 14$) fillings for $V>V_c$.
At the commensurate fillings, 
$O_\Psi$ first exhibits a sharp decrease and then, 
it returns to the original value quite rapidly as $V$ increases.
On the other hand at incommensurate fillings,
this recurrence of $O_\Psi$ does not occur.
These results mean that the GS evolves into a novel localized state
at commensurate filling,
whereas at the incommensurate fillings, the GS moves to an extended state.
This observation is obviously in good agreement with the OPDM in Fig.~\ref{Fig2}.

The system-size dependence is also displayed in Fig.~\ref{Fig3}.
For the case with tiny off-diagonal elements of the OPDM,
the lower bound of $O_\Psi$ is estimated as $\sim N^2/2L$, and therefore 
$O_\Psi/ 2L \sim (N/2L)^2=1/4$ in the half-filling case.
Calculations show slightly larger value than $1/4$, indicating the existence of non-vanishing off-diagonal
elements in the OPDM even in the vicinity of the transition as can be seen in Fig.~\ref{Fig2} for $V=5$.
As the system size is getting larger, the range of $V$ for the sharp decrease of 
$O_\Psi$ is getting smaller,
indicating the possibility that this behavior persists in the limit 
$L\rightarrow \infty$.
[Increase of the minimum of $O_\Psi$ as $L$ increases comes from the fact
that the dimension of the OPDM gets larger for larger $L$,
and it does not imply that the transition gets weak for larger $L$.]

%%%%%%%%%%%%%%%%%%%%%%%%%%%%%%%%%%%% 
\begin{figure}[t]
\begin{center} 
\includegraphics[width=8cm]{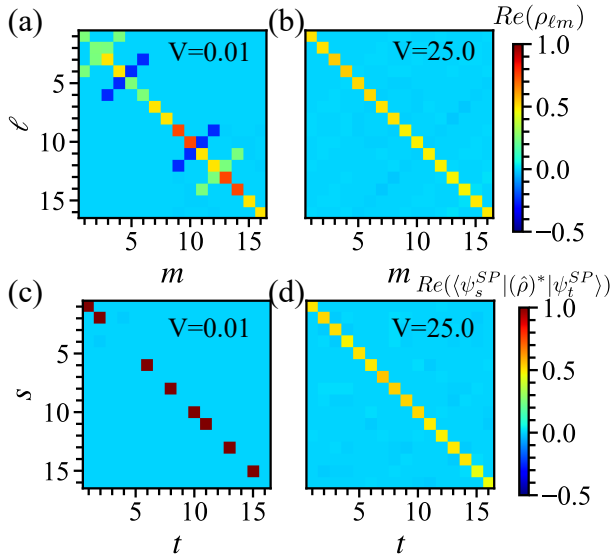}  
\end{center} 
%\vspace{-0.5cm}
\caption{
(a) and (b): Real part of OPDM of an eigenstate in the middle of the spectrum 
for $V=0.01$ and $25.0$.
The OPDM is featureless for large $V$.
(c) and (d): Real part of $\langle \psi^{SP}_{s}|(\hat{\rho})^{*}|\psi^{SP}_{t}\rangle$. 
The state labels $s$ and $t$ are in the ascendant energy order.
Each result is obtained from single shot for single disorder realization. 
}
\label{Fig4}
\end{figure}
%%%%%%%%%%%%%%%%%%%%%%%%%%%%%%%%%%%%
%%%%%%%%%%%%%%%%%%%%%%%%%%%%%%%%%%% 
\begin{figure}[t]
\begin{center} 
\includegraphics[width=5.5cm]{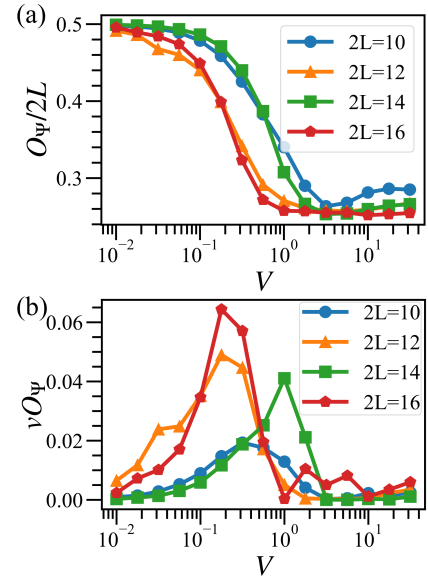}  
\end{center} 
\caption{
(a)$O_\Psi=\langle\mbox{Tr}  [\hat{\rho}^2]\rangle$ of energy-spectrum-middle states. 
$O_\Psi$ decreases in all cases, but exhibits a difference behavior in
the commensurate ($2L=12, 16$) and incommensurate ($2L=10, 14$) fillings 
as in the GS shown in Fig.~\ref{Fig3}.
This implies that the correlations of the rung density wave play some role
in the instability of localization.
(b)Variance of $O_\Psi$, $vO_\Psi$, for various system sizes.
The results indicate a second-order transition of energy-spectrum-middle states.
}
\label{Fig5}
\end{figure}
%%%%%%%%%%%%%%%%%%%%%%%%%%%%%%%%%%%%

Figures~\ref{Fig4} (a) and ~\ref{Fig4} (b) show the real part of the OPDM of an eigenstate
in the middle of the spectrum (nearly zero energy).
For small $V$, the state is mainly described by a product of a pair of 
$(W^\pm_j)^{\dagger}|0\rangle$ Fock state. 
As $V$ is getting large, almost all states in the Fock-space take part in many-body states.
Also, the real part of the OPDM in the $W$-basis
are shown in Figs.~\ref{Fig4}(c) and \ref{Fig4}(d).
The all off-diagonal elements of the real-space OPDM are almost vanishing,
and from perspective of the Fock-space, this means that the coefficients of the expansion
such as $|E\r=\sum_IA_I|I\r$ take random variables and as a result, 
$\rho_{\ell m}\sim 0 \ (\ell\neq m)$. 
This explains the result that the real part of the off-diagonal elements of 
$\langle \psi^{SP}_{s}|(\hat{\rho})^{*}|\psi^{SP}_{t}\rangle$ are also small. 
In addition, $O_\Psi$ for the middle of the spectrum is shown in
Fig.~\ref{Fig5}(a), and it decreases rapidly as $V$ increases and keeps very 
small values for larger $V$ in contrast to the GS.
The above observation 
$\rho_{\ell m}\sim 0 \ (\ell\neq m)$ is supported by the calculation of $O_\Psi$
showing $O_\Psi /2L\sim 1/4$ for $V>1$.
The localized states in the middle of the energy-spectrum are destroyed by the repulsion.

In Fig.~\ref{Fig5}(a), we also show the system-size dependence of 
$O_\Psi$ for the spectrum-middle states.
The calculation indicates that the change from the localized phase to the extended state 
is a second-order transition as $O_\Psi$ exhibits clear system-size dependence.
The result also implies that the correlations of the rung density wave play some role
in the instability of localization as in the GS in Fig.~\ref{Fig3}.
Discussion on solvable models in Sec.~IV will shed light on this behavior.
It is also useful to measure the variance of
$O_\Psi$, $vO_\Psi \equiv {1 \over 2L}\langle (\mbox{Tr}[\hat{\rho}^2]- O_\Psi)^2\rangle$,
calculation of which is displayed in Fig.~\ref{Fig5}(b).
The system-size dependence of $vO_\Psi$ obviously supports the above conclusion of 
the state transition
of the spectrum-middle regime.

%%%%%%%%%%%%%%%%%%%%%%%%%%%%%%%%%%%% 
\begin{figure}[t]
\begin{center} 
\includegraphics[width=8.5cm]{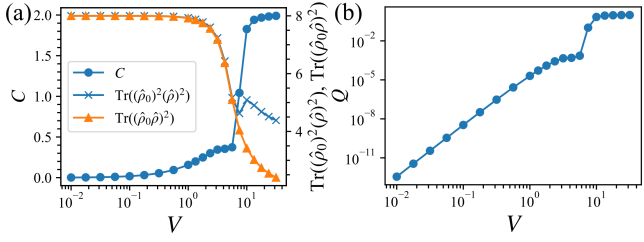}  
\end{center} 
%\vspace{-0.5cm}
\caption{
(a) Frobenius norm ${\cal C}$ of the GS.
We also plot $\mbox{Tr}[(\hat{\rho}_0)^2(\hat{\rho})^2]$, and
$\mbox{Tr}(\hat{\rho}_0\hat{\rho})^2$.
(b) ${\cal Q}$ is a smoothly increasing function of $V$ till $V\simeq 7$.
Step-wise increase indicates a phase transition to another localized state.
Averaged over $5000$ samples of the eigenstates.
}
\label{Fig6}
\end{figure}
%%%%%%%%%%%%%%%%%%%%%%%%%%%%%%%%%%%%

The above observation of the OPDM indicates the importance of the magnitude of the off-diagonal
elements to search state transition induced by the interactions.
To quantify the behavior of the GS as a function of $V$, we measure quantities 
such as \cite{Chen2021}
\be
{\cal C}\equiv\sqrt{\mbox{Tr}[[\hat{\rho}_0, \hat{\rho}][\hat{\rho},\hat{\rho}_0]]}
=\sqrt{2\mbox{Tr}[(\hat{\rho}_0)^2(\hat{\rho})^2-(\hat{\rho}_0\hat{\rho})^2]},
\nonumber
\ee
where
$\hat{\rho}=\{\langle \psi^{SP}_{s}|(\hat{\rho})^{*}|\psi^{SP}_{t}\rangle$\}
is the $W$-basis OPDM for finite $V$
and $\hat{\rho}_0$ is the $W$-basis OPDM for $V=0$,
and we also measure
\be
{\cal Q}\equiv \sum |\mbox{off-diagonal elements of} \ \hat{\rho} |^4.
\nonumber
\ee
The quantity ${\cal C}$ is called Frobenius norm, and it quantifies the distance 
between $\hat{\rho}_0$ and $\hat{\rho}$.
In Fig.~\ref{Fig6}, we display the numerical results of these quantities for the GS.
Both ${\cal C}$ and ${\cal Q}$ exhibit a step-wise change at $V\simeq 7$ indicating 
the existence of a state transition.
In particular, 
the Frobenius norm can be used as an order parameter for observing
that the GS enters into another localized state described by the Slater determinant composed of
eigenstates of the OPDM in Fig.~\ref{Fig2}(c).
To investigate the physical picture of the newly emergent state, study on a solvable model
is quite useful, as we see in the subsequent section.

%%%%%%%%%%%%%%%%%%%%%%%%%%%%%%%%%%%%%%%%%%%%%%%%%%%%
\section{Study by solvable models}

In the preceding section, we investigated the 
phase transition of $H_{\cal T}$ at half-filling by using the numerical calculation.
In this section, we shall study some related interacting model, which is exactly solvable due to the existence 
of local integrals of motion (LIOMs) \cite{OKI2020}.
Hamiltonian of the target model is given as follows: $H_{\cal P}=H_{\rm rh}+H_{\rm N}$,
\be
H_{\rm N}=g\sum^{L}_{j=1}N_{j-1}[W^{+\dagger}_jW^-_j+W^{-\dagger}_jW^+_j],
\label{HN}
\ee
where $N_j\equiv N^+_j+N^-_j$, $N^\pm_j=W^{\pm\dagger}_jW^\pm_j$, and $g$ is a coupling constant.
$H_{\cal P}$ is a projective Hamiltonian.
To show this, we display the following LIOMs, $K^\pm_j \ (j=1,\cdots, L)$,
\be
&& K^+_j \equiv N^+_j-{g \over 4t_j} O_j,  \;\;
K^-_j \equiv N^-_j+{g \over 4t_j} O_j.\nonumber
\ee
where $O_j = N_{j-1}[W^{+\dagger}_jW^-_j+W^{-\dagger}_jW^+_j]$.
It is verified that $\{K^{\pm}_j\}$'s commute with each other, and $H_{\cal P}$ is expressed as,
$H_{\cal P}=\sum^{L}_{j=1}2t_j(-K^+_j+K^-_j)$.

As shown in the above, $H_{\cal P}$ is a projective Hamiltonian, and
all eigenstates of $H_{\cal P}$ are given by eigenstates of $\{ K^\pm_j\}$'s.
To see how eigenstates look like, let us focus on states at plaquette $j$ and study eigenstates 
and eigenvalues for the case of $\langle N_{j-1}\rangle=1$.
We put $\lm_j=g\l N_{j-1}\r/(4t_j)$, then after some manipulation, we obtain eigenstates, $\psi_1$ and $\psi_2$ as
\be
&& \psi_1\propto (x_{1j} W^{+\dagger}_j-\lambda_j W^{-\dagger}_j)|0\rangle, \nonumber  \\
&& \psi_2 \propto (\lambda_j  W_j^{+\dagger}+x_{1j} W_j^{-\dagger})|0\rangle. 
\label{waveF12}
\ee
These eigenstates satisfy the followings: 
\be
&&K^+_j \psi_1=x_{1j}\psi_1, \; \;K^+_j\psi_2=x_{2j}\psi_2,  \nonumber  \\
&&K^-_j \psi_1=x_{2j}\psi_1, \;\; K^-_j\psi_2=x_{1j}\psi_2,
\nonumber
\ee
where $x_{1j}={1 \over 2}(1+\sqrt{1+4\lambda_j^2})$ and $x_{2j}={1 \over 2}(1-\sqrt{1+4\lambda_j^2})$.
The GS of $H_{\cal P}$ at half-filling is given by 
$|{\rm GS}\rangle= \prod_{j=1}^L\psi_{1j}$ with $\langle N_j\rangle=1$ for all $j$, and energy, 
$E_{\rm GS}=-2\sum_jt_j(1+4\lm^2_j)^{1/2}.$
Deformation of the state by the interactions is easily seen as it changes $W^{+\dagger}_j|0\r \to \psi_{1j}$, i.e.,
\textit{
uniformly distributed fermion density is deformed by the interactions, $H_{\rm N}$.}
More precisely, the element at the $j$-rung, $(a_j-ib_j)$, gets larger amplitude than that at the $(j+1)$ rung,
$(a_{j+1}+ib_{j+1})$ \cite{psi1_detail}.
The OPDM of the GS has diagonal elements $\trho_{jj}=x^{2}_{1j}/(x^{2}_{1j}+\lm_j^2)$ for $i\leq N$
and $\trho_{jj}=\lm_j^2/(x^{2}_{1j}+\lm_j^2)$ for $i >N$.
The anti-diagonal elements are also non-vanishing as
$\l W^{+\dagger}_jW^-_j\r=-\lm_j x_{1j}/(x^2_{1j}+\lm_j^2)$.

%%%%%%%%%%%%%%%%%%%%%%%%%%%%%%%%%%%% 
\begin{figure}[t]
\begin{center} 
\includegraphics[width=6cm]{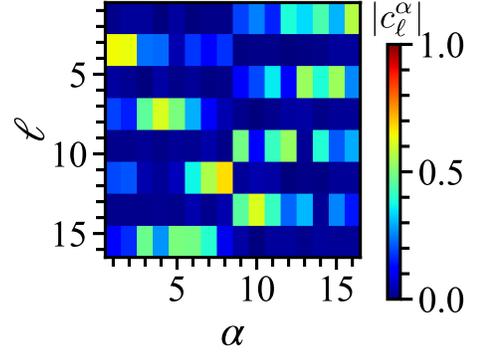}  
\end{center} 
%\vspace{-0.5cm}
\caption{
Density of eigenvectors of the real-space OPDM for $V=30$.
$\alpha$ labels eigenvectors and $j$ lattice sites (for the original fermion $c^{\dagger}_\ell$, $\ell=1,2,\cdots, 2L$).
$\{c_\ell^\alpha\}$ are expansion coefficients.
Each eigenvector resides on a rung, as expected by the analytical study on the solvable model, $H_{\cal P}$.
}
\label{Fig7}
\end{figure}
%%%%%%%%%%%%%%%%%%%%%%%%%%%%%%%%%%%%

We recognize similar behavior to the above analytical results in the numerical calculations of the OPDM for $H_{\cal T}$.
[Please see Figs.~\ref{Fig2}(e) and ~\ref{Fig2}(f), and recognize that the mixing amplitudes, $\l W^{+\dagger}_jW^-_j\r$, 
take both positive and negative values.]
In fact, this is not pure coincidence.

It was discussed and verified by the practical calculations that 
the leading terms of uniform interactions such as $H_{\rm I}$ in Eq.~(\ref{HI1}) are 
those described solely
in terms of the LIOMs of the system without the interactions~\cite{Pollmann2019},  
and in the system $H_{\cal T}$, these LIOMs are nothing but density operators $\{N^\pm_j\}$.
These leading terms do not deform the wave functions of the energy eigenstates, 
as they are
simultaneous eigenstates of  $\{N^\pm_j\}$.
The next-leading terms out of $H_{\rm I}$, $H^I_N$, include flipping terms of $W^\pm_j$ besides the LIOMs,
and are obtained by the practical calculation as follows;
\be
H^I_{\rm N}&=&{V \over 2}\sum^{L}_{j=1}[N_{j+2}+N_{j+1}-N_{j-2}-N_{j-1}]  \nonumber  \\
&& \times [W^{+\dagger}_jW^-_j+W^{-\dagger}_jW^+_j],
\label{HNI}
\ee
which have similar structure to $H_{\rm N}$ in Eq.~(\ref{HN}) and they construct a projective Hamiltonian.
Therefore, we expect that the leading-order repulsion described solely by $\{N^\pm_j\}$ induces inter-plaquette 
density fluctuations and $H^I_{\rm N}$ [Eq.~(\ref{HNI})] induces the $(W^+-W^-)$ mixing
and as a result, intra-plaquette density fluctuations emerge.
In order to verify this observation, we show eigenvectors of the real-space OPDM, $\hat{\rho}$, 
for $V=30$ [displayed in Fig.~\ref{Fig2}(c)] in Fig.~\ref{Fig7}.
\textit{Each eigenvector resides on a rung}, as predicted by the analytical study on 
the solvable model, $H_{\cal P}$.
By the above consideration, we can write down the (approximate) wavefunction of the GS 
of the Hamiltonian $H_{\cal T}$ for the strong interactions:
$|GS_\infty\r=\prod_{j=0}^{{L\over 2}-1}[\psi^-_{2j+1}\psi^+_{2j+2}]$,
where $\psi^{+(-)}_j = \psi_{1j}|_{g \to \pm\infty}$.
[Please note that the GS has the rung-density-wave order~\cite{FN1}.]

One may wonder if binary LIOMs can be constructed from $\{K^\pm_j\}$, whose eigenvalues 
are $1$ or $0$.
We comment on this point.
In fact, it is feasible, and they are explicitly given by
$n_{\gamma j}=\psi^\dagger_{\gamma j}\psi_{\gamma j}$ ($\gamma=1,2$) with
`quasi-particle' operators;
$
\hat{\psi}^\dagger_{1j}={1 \over \sqrt{\A}}\
\Big(\hat{x}_jW^{+\dagger}_j-\hat{\lambda}_jW^{-\dagger}_j\Big),
$ 
and 
$
\hat{\psi}^\dagger_{2j}={1 \over \sqrt{\A}}\
\Big(\hat{\lambda}_jW^{+\dagger}_j+\hat{x}_jW^{-\dagger}_j\Big),
$
where we have defined \textit{operators} $\hat{\lm}_j=g\hat{N}_{j-1}/(4t_j)$, etc.
Details of these operators are scrutinized in Appendix B.
In the text, we shall explain essential aspects of these operators.
We can prove $\{\hat{\psi}^\dagger_{\gamma j}, \hat{\psi}_{\gamma' j}\}=\delta_{\gamma \gamma'}$.
However, as the operation of $\hat{\psi}_{1(2)j}$ changes the number of fermions residing in the plaquette $j$,
they do \textit{not} anti-commute with $\hat{\psi}_{\gamma j+1}$ ($\hat{\psi}^\dagger_{\gamma j+1}$), 
and therefore cautious manipulation with them is required. 
However, their number operators, $n_{\gamma j}$, commute with $N_j=N^+_j+N^-_j$, and therefore,
in the sector of the Hilbert space with fixed value of $N_j$, the operators $\hat{x}_j=x_{1j}$
and $\hat{\lambda}_j=\lambda_j$ can be treated as c-number.
Hamiltonian, $H_{\cal P}$, is expressed as 
\be
\tilde{H}_{\cal P}= -\sum^{L}_{j=1}2t_j \sqrt{1+4\hat{\lambda}^2_j}\ \Big[\psi^\dagger_{1j}\psi_{1j}-\psi^\dagger_{2j}\psi_{2j}\Big],
\label{HP2}
\ee
and for each sector by substituting corresponding c-number for  $\hat{x}_j$ and $\hat{\lambda}_j$, 
energy eigenvalues and eigenfunctions are obtained by solving 
the system $\tilde{H}_{\cal P}$.
If we expand $\tilde{H}_{\cal P}$ in Eq.~(\ref{HP2}) in power of $g$,
we obtain a reminiscence of the effective Hamiltonian of localization
expressed in terms of the LIOMs.

%\bigskip

%%%%%%%%%%%%%%%%%%%%%%%%%%%%%%%%%%%%%%%%%%%%%%%%%%%%%%%%%%%%%%
\section{Discussion and conclusion}

In this article, we studied effects of the repulsion to the random-hopping Creutz ladder. 
We mostly employed the OPDM for clarifying the phase diagram, localization transition, 
and structure of the GS and spectrum-middle states. 
The OPDM elucidates how interactions deform many-body states composed of 
compact localized states.
The OPDMs expressed by the real-space basis as well as the compact-localized state basis
are complementary with each other and various quantities are extracted from them by 
using the numerical
calculations.
Study of a solvable model with the explicit LIOMs also gives useful insights to 
understand the detailed structure of numerically obtained many-body states in 
the random-hopping Creutz ladder. 
We obtained clear pictures of the phases concerning the GS and states in the middle of 
the spectrum.
We showed that the OPDM contains rich information about localized states and 
localization transition.
It is a future work to apply the obtained utilities to other models.

Finally, we comments on the properties of the GS transition observed by 
the various quantities.
Eigenvalues of the OPDM in Fig.~\ref{Fig1}, as well as $O_\Psi$, ${\cal C}$ and 
${\cal Q}$ exhibit
a step-wise discontinuity at the transition point $V_c \simeq 7$, indicating 
that the GS transition is of first order. 
At the first-order phase transition, the correlation length (of the rung density
wave) can be finite and not diverge.
This observation does not contradict the recent study~\cite{Sahay2021}
indicating the existence of an intervening
ETH phase between two distinct MBL states, as we are studying the GS.
In general, to extract the trustworthy localization length in MBL is not an easy task, but methods to use the participation
ratio obtained by the OPDM was proposed and it seems to work successively~\cite{Bera2015}.
For the present system, however, the method is difficult to be applied because of 
the degeneracy of the many-body states. Therefore, a definite answer to the question 
if the localization length diverges or not at the critical point cannot be obtained, 
even though the correlation length might be finite by the properties of 
the first-order transition.
However, qualitative argument may be possible.
That is, we understand that at the transition point, the GS changes from the state of the plaquette-order
to the rung-order.
In order for this transition to take place, emergence of an extended state is not needed as the change of the 
GS occurs locally. Anyway, to answer to the above question is definitely an
important and interesting future problem.

There are a few interesting works studying a transition between MBL states in disordered spin 
chains~\cite{Pekker2014,Kajall2014}.
These studies showed that the MBL regime is divided into two MBL phases, i.e., paramagnetic and spin-glass
MBL phases.
Behavior of domain walls is different in the two phases, i.e., localized domain walls are created and removed
in pairs in the paramagnetic MBL state, whereas they are stable without overlapping each other in 
the spin-glass MBL state.
For the GS of the system studied in this work, we also observed that 
the spatial characteristics in each of two phases, i.e., the plaquette and 
rung-order density and also phase-coherence pattern, change.
This observation of the similarity might be useful for future investigation of phase transitions between 
multiple MBL phases.

\bigskip

\textit{Acknowledgements.---}
T.O. has been supported by the Program for Developing and Supporting the Next-Generation of Innovative Researchers at Hiroshima University. This work is also supported by JSPS KAKEN-HI Grant Number JP21K13849 (Y.K.).
%\clearpage 

\appendix
%%%%%%%%%%%%%%%%%%%%%%%%%%%%
\section*{Appendix}
%%%%%%%%%%%%%%%%%%%%%%%%%%%%%%%%%%%%%%%%%%%%%%%%%
\subsection*{A. OPDM in critical regime}
%%%%%%%%%%%%%%%%%%%%%%%%%%%%%%%%%%%% 
\begin{figure}[h]
\begin{center} 
\includegraphics[width=8cm]{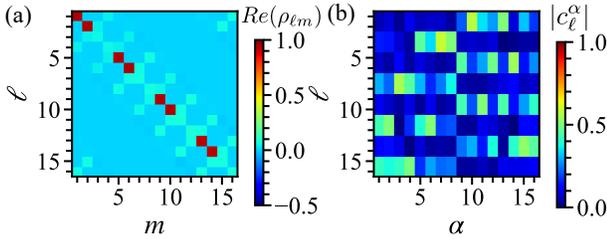}  
\end{center} 
\caption{
(a) OPDM for ground state in the representation in the real space basis.
Rung density wave with small $W^\pm$-type fluctuations is observed.
(b) Density of eigenvectors of the real-space OPDM, $\rho$ [Eq.(2) in the main text] for $V=10$.
Each eigenvector resides on a rung, as expected by the analytical study on the solvable model, $H_{\cal P}$.
}
\label{Fig8}
\end{figure}
%%%%%%%%%%%%%%%%%%%%%%%%%%%%%%%%%%%%

In the main text, we showed the OPDM for small, intermediate and large values of $V$.
In this Appendix, we display similar data for the ground state of the system 
just after the phase transition, i.e., $V=10$.
In the OPDM in the real-space basis [Fig.~\ref{Fig8}(a)], rung density wave with small $W^\pm$-type fluctuations 
is observed. 
In Fig.~\ref{Fig8}(b), eigenvectors of the OPDM are displayed, and almost all of them seem
to reside on a rung forming rung density wave, although there are fluctuations from the exact density wave.
The above results for the $V=10$ system are in good agreement with data of $O_\Psi$, ${\cal Q}$ and ${\cal C}$
indicating the phase-transition value $V_c\simeq 7$.

%%%%%%%%%%%%%%%%%%%%%%%%%%%%%%%%%%%%%%%%%%%%%%%%%%%%
\subsection*{B. Quasi-particle operators}
In this Appendix, we shall discuss the solvable model, $H_{\cal P}$,  a little bit more,
in particular, `quasi-particle' operators.
In the main text, we obtained eigenstates of the LIOMs, $K^\pm_j$, as follows:
\be
&&\psi_{1j}\propto (x_{1j} W^{+\dagger}_j-\lambda_j W^{-\dagger}_j)|0\rangle, \nonumber  \\
&&\psi_{2j} \propto (\lambda_j  W_j^{+\dagger}+x_{1j}W_j^{-\dagger})|0\rangle.
\label{QP1}
\ee
Equations~(\ref{QP1}) indicate that one may obtain quasi-particle operators as,
\be
\hat{\psi}^\dagger_{1j}&=&{1 \over \sqrt{\A}}\
\Big(\hat{x}_jW^{+\dagger}_j-\hat{\lambda}_jW^{-\dagger}_j\Big),   \nonumber  \\
\hat{\psi}^\dagger_{2j}&=&{1 \over \sqrt{\A}}\
\Big(\hat{\lambda}_jW^{+\dagger}_j+\hat{x}_jW^{-\dagger}_j\Big).
\label{QP2}
\ee
In fact, the operators $\hat{\psi}^\dagger_{1j}$ and  $\hat{\psi}^\dagger_{2j}$ produce the states
in Eq.~(\ref{QP1}) correctly when \textit{they are applied to the empty state at plaquette $j$}, i.e.,
$\hat{\psi}^\dagger_{1j}|0\r = \psi_{1j}/|\psi_{1j}|$.
However, $\hat{\psi}^\dagger_{\gamma j}$ ($\gamma=1,2$) does not anti-commute with 
$\hat{\psi}^\dagger_{\gamma', j-1}$ nor $\hat{\psi}_{\gamma', j-1}$ as $\hat{x}_j$ and $\hat{\lambda}_j$ 
contain the number operator at $(j-1)$, $\hat{N}_{j-1}$. 
Therefore, $\hat{\psi}^\dagger_{1j}$ and  $\hat{\psi}^\dagger_{2j}$ are \textit{not} quasi-particle operators.

However, their number operators, $n_{\gamma j}=\hat{\psi}^\dagger_{\gamma j}\hat{\psi}_{\gamma j}$
commute with $\hat{N}_{j-1}$ and themselves.
Practical calculation shows,
\be
\psi^\dagger_{1j}\psi_{1j} &=& {1 \over \B} \ \Big[x^2_jW^{+\dagger}_jW^+_j+\lambda^2_jW^{-\dagger}W^-_j
\nonumber \\
&&-x_j\lambda_j(W^{+\dagger}_jW^-_j+W^{-\dagger}_jW^-_j)\Big],   \nonumber \\
\psi^\dagger_{2j}\psi_{2j} &=& {1 \over \B} \ \Big[\lambda^2_jW^{+\dagger}_jW^+_j+x^2_jW^{-\dagger}W^-_j
\nonumber \\
&&+x_j\lambda_j(W^{+\dagger}_jW^-_j+W^{-\dagger}_jW^-_j)\Big], 
\label{numberpsi}
\ee
where we have set $\hat{x}_j (\hat{\lambda}_j) \to x_j (\ \lambda_j)$ as they commute with $N_{j'}$.
From Eq.~(\ref{numberpsi}), we have
\be
\psi^\dagger_{1j}\psi_{1j}-\psi^\dagger_{2j}\psi_{2j}
&=&-{x_j\over \B}\Big[-W^{+\dagger}_jW^+_j+W^{-\dagger}_jW^-_j  \nonumber  \\
&&+{gN_{j-1} \over 2t_j}(W^{+\dagger}_jW^-_j+W^{-\dagger}_jW^+_j)\Big] \nonumber \\
&=&-{x_j\over \B}[-K^+_j+K^-_j].  
\label{numberpsi2}  
\ee
\be
(\psi^\dagger_{1j}\psi_{1j}+\psi^\dagger_{2j}\psi_{2j}
&=& W^{+\dagger}_jW^+_j+W^{-\dagger}_jW^-_j,\nonumber\\ 
&&\psi^\dagger_{1j}\psi^\dagger_{2j} = W^{+\dagger}_jW^{-\dagger}_j )
\label{numberpsi3} 
\ee
Therefore, Hamiltonian $H_{\cal P}$ is expressed as,
\be
\tilde{H}_{\cal P}= -\sum^{L}_{j=1}2t_j \sqrt{1+4\hat{\lambda}^2_j}\ \Big[\psi^\dagger_{1j}\psi_{1j}-\psi^\dagger_{2j}\psi_{2j}\Big].
\label{Hpsi}
\ee
It should be emphasized that there exist interactions between the number operators,
$n_{\gamma j}$, in $\tilde{H}_{\cal P}$ [Eq.~(\ref{Hpsi})]
as $\{\hat{\lambda}_j\}$ contain $\{n_{\gamma, j-1}\}$. 
One may expect that the usual logarithmic increase in entanglement entropy (EE)
takes place in a global quench of this system.
However, the system's Hamiltonian is expressed as, 
$$
H_{\cal P}=\sum^{L}_{j=1}2t_j(-K^+_j+K^-_j)
$$ by the LIOMs, $K^\pm_j$,
and therefore EE may exhibit more stable behavior
in a global quench.
In fact, our recent work on some related spin model~\cite{KOIspin} shows 
that it exhibits stable EE similar to that in Anderson localization.

%%%%%%%%%%%%%%%%%%%%%%%%%%%%%%%%%%%%%%%%%%%%%%%%%
%%%%%%%%%%%%%%%%%%%%%%%%%%%%%%%%%%%%%%%%%%%%%%%%%%%%%%%%%%%%%%%%%%%%%%%%%%%%%%%


\begin{thebibliography}{99}
\bibitem{psi1_detail}
The explicit form of $\psi_{1}$ is
$\psi_1=\frac{1}{2}(x_{1j}-\lambda_{j})(ia^{\dagger}_{j+1}+b^{\dagger}_{j+1})|0\rangle
+\frac{1}{2}(x_{1j}+\lambda_{j})(a^{\dagger}_{j}+ib^{\dagger}_{j})|0\rangle$. 
Due to $\lambda_j>0$, the amplitude of $(a^{\dagger}_{j}+ib^{\dagger}_{j})$ is larger 
than that of $(ia^{\dagger}_{j+1}+b^{\dagger}_{j+1})$.
\bibitem{FN1}
Data in Fig.~\ref{Fig7} show small but finite deviations of particle density
on rungs from the perfect rung density wave.
We think that this comes from the mixing terms of $W^\pm$ located 
on the next-nearest-neighbor plaquettes such as 
$(N^+_{j}-N^-_j)(W^{-\dagger}_{j-1}W^+_{j+1}
-W^{+\dagger}_{j-1}W^-_{j+1} +\mbox{h.c.})$, which emerge from
$H_{\rm I}$, besides those in Eq.~(\ref{HNI}).
\end{thebibliography}
\end{document}